\documentstyle[psfig]{aipproc}
\begin{document}
\def\Sp{\mbox{Sp}}
\def\erfc{\mbox{erfc}}
\def\Re{\mbox{Re}}
\def\Im{\mbox{Im}}
\title{Dynamical symmetry breaking in Nambu-Jona-Lasinio model under the
influence of external electromagnetic and gravitational fields}
\author{E. Elizalde$^{a,b,}$\thanks{E-mail: eli@zeta.ecm.ub.es,
elizalde@io.ieec.fcr.es}, 
Yu. I. Shil'nov$^{a,c,}$\thanks{E-mail: shil@kink.univer.kharkov.ua,
 visit2@ieec.fcr.es} \\
$^a${\it Consejo Superior de Investigaciones Cient\'{\i}ficas,}\\
{\it IEEC, Edifici Nexus-204, Gran Capit\`a 2-4, 08034, Barcelona, Spain}\\
$^b${\it Department  ECM, Faculty of Physics, University of Barcelona,}\\
{\it Diagonal 647, 08028, Barcelona, Spain}\\
$^c$ {\it Department of 
Theoretical Physics, Faculty of Physics,}\\
{\it Kharkov State University, Svobody Sq. 4, 310077,
Kharkov, Ukraine}
}
\maketitle
\begin{abstract}
Dynamical symmetry breaking  is investigated for a four-fermion
Nambu-Jona-Lasinio model in external electromagnetic and gravitational 
fields. An effective 
potential is calculated in the leading order of the large-N expansion using
the proper-time Schwinger formalism. 

Phase transitions accompanying a chiral symmetry 
breaking in the Nambu-Jona-Lasinio model are studied in detail. 
A magnetic calalysis phenomenon is shown to exist in  curved spacetime 
but it turns out to lose its universal character because
the chiral symmetry is restored  above some  
critical positive  value of the spacetime curvature.
\end{abstract}
\section*{Introduction}

Different four-fermion models \cite{ESH:NJL}, \cite{ESH:GN} have 
been considered to be one of 
the most convenient ways for an investigation of the low-energy physics of 
strong interactions.
A dynamical symmetry breaking phenomenon (DSB) has been proved to take place  
within those models, partecularly Nambu-Jona-Lasinio (NJL) one, which 
seems to show up a nontrivial phase structure. 
Usually the symmetry to be broken
under the DSB mechanism is the chiral one. Dynamical version of fermions
mass generation and dynamical chiral symmetry breaking have been
investigated very carefully and some fruitful applications   
for the real high-energy physics have been found \cite{ESH:DSB}, 
\cite{ESH:PhysRep}.
\pagebreak

However it has turned out to be very difficult to realize the idea of 
DSB because all of the calculations should be performed out 
of  pertubation theory. This leads to study  already simplified models
and that is why we have to 
investigate  any possible 
generalizations within these models those of nonzero temperature and chemical 
potential, arbitrary dimensions, external fields including gravitational 
one and so on as  some kind of laboratory in order to collect as much new 
information as we can.

Despite of essential difficulties caused by the nonpertubative character of  DSB
 phenomenon, it has been 
applied successfully to describe 
the overcritical behavior of quantum electrodynamics, the top quark condensate
mechanism of mass generation in the Weinberg- Salam model of electroweak
interactions,  technicolor models and, especially, to investigate
the composite fields generation in the NJL model.
In the frameworks of Schwinger proper-time method
this model has been 
studied in external electromagnetic field 
by many authors for 20 years  \cite{ESH:Sch} - \cite{ESH:emf}.

Recently  a new sample of papers devoted to DSB in external electromagnetic
field have been 
published \cite{ESH:emf2}, \cite{ESH:GMS}. 
It shed a new light onto the universal character of magnetic catalysis,
which  means 
that magnetic field breaks  chiral symmetry for any value of 
its strength. Furthermore it has been shown that this phenomenon occurs
in quantum electrodynamics, 2+1, 3+1 dimensional nonsupersymmetrical and 
3+1 supersymmetrical NJL
models. So the statement about {\bf the universal character of magnetic catalysis}
has been made.
   
Investigations of the influence of  a classical gravitational field on
the DSB phenomenon in the NJL model have been carried out for some years
\cite{ESH:ELOS}.
It has been shown that curvature-induced phase transitions exist and might 
play some essential role in more or less realistic early Universe model.
It turns out that, in spite of the relatively small value of the
 curvature-dependent
corrections at the low energy scale to be investigated within
the NJL model, these corrections appear to be inescapable, in the sense
that they must be taken into account when one performs
the necessary "fine tuning" of the different cosmological parameters.
Furthemore, positive spacetime curvature changes the universal
character of magnetic catalysis dramatically.

It has been shown that the early Universe could contain
a large primodial magnetic field and have a huge electrical conductivity. 
The vicinities of magnetized 
black holes and neutron stars are the other possible
points of application of our model. 
Therefore  both classical external gravitational
and electromagnetic fields should be taken into account for 
the description of a wide sample of events in the Universe. 

In the present paper we describe our recent results concerning the DSB
under the simultaneous influence of both gravitational and electromagnetic 
fields in NJL model \cite{ESH:OWN}. The phase transitions accompanying the 
DSB process
on the spacetime curvature, as well as the values of electric or magnetic 
field strength are investigated.
\newpage  
\section*{Dynamical symmetry breaking by a magnetic field in flat spacetime}

In an arbitrary dimensional flat spacetime the NJL model has the 
following action:
\begin{equation}
 S=\!\int d^d x \left\{ i \overline{\psi}\gamma^\mu D_\mu \psi +
{\lambda \over 2N} \left[ (\overline{\psi}\psi)^2+
(\overline{\psi} i \gamma_5 \psi)^2 \right] \right\},
\end{equation}
where the covariant derivative $D_{\mu}$ includes the electromagnetic
potential $A_{\mu}$ and  $N$ is the number of bispinor fields $\psi_a$.

Introducing the auxiliary fields
\begin{equation}
\sigma=-{\lambda \over N }(\overline{\psi} \psi ), 
\pi=-{\lambda\over N}( \overline{\psi} i \gamma_5 \psi)
\end{equation}
we can rewrite the action as:
\begin{equation}
S=\int d^d\! x  \left\{ i\overline{\psi}\gamma^\mu D_\mu \psi -
{N \over 2\lambda}(\sigma^2+\pi^2)-
\overline{\psi}(\sigma+i\pi\gamma_5)\psi \right\}.
\end{equation}

The effective action in the leading $1/N$ order is
\begin{equation}
\frac{1}{N} S_{eff}=-\int d^d\! x {\sigma^2+\pi^2 \over 2\lambda}-
i \ln \det\left[ i\gamma^\mu D_\mu-(\sigma+i\gamma_5\pi)\right]
\end{equation}

Then the effective potential (EP), 
defined for the constant configurations of
$\pi$ and $\sigma$ as 
$V_{eff} = -S_{eff}/ N\! {\displaystyle \int}\! d^d\!x$, is given by 
the formula 
\begin{equation}
V_{eff}={\sigma^2 \over 2\lambda }+i \Sp \ln \langle x| [ i\gamma^\mu D_\mu - 
\sigma ] |x \rangle 
\end{equation}
Here we put $\pi=0$ because the  final expression will depend on the combination 
$\sigma^2+ \pi^2$ only within our approximation. This means  that we are actually
considering  the Gross-Neveu model. 

But if we take into account kinetic terms of the fields $\pi$ and $\sigma$ 
generated by quantum corrections we will obtain  different dynamics of these 
two  fields. It should be noted that $\sigma$ will be a massive scalar field
in the supercritical area 
while $\pi$ will be massless Goldstone particle.

By means of the usual Green function (GF), which obeys the equation
\begin{equation}
(i \gamma^\mu D_\mu-\sigma)_x G(x,x',\sigma)=\delta(x-x'),
\end{equation}
we obtain the following formula
\begin{equation}
V_{eff}'(\sigma)={ \sigma \over \lambda }-i \Sp G(x,x,\sigma).
\end{equation}

Now we can substitute in this equation the
fermion GF in constant magnetic field
\begin{equation}
G(x,x',\sigma)=\Phi(x,x')\tilde{G}(x-x',\sigma),
\end{equation}
where
\begin{equation}
\Phi(x,x')= \exp  \biggl[ ie\int\limits^x_{x'} A^\mu (x'') dx'' \biggr]
\end{equation}
\begin{eqnarray}
\tilde{G_0}(z,\sigma)=e^{-i{\pi\over 4}d}\int\limits_0^\infty
\frac{ds}{(4\pi s)^\frac{d}{2}}
e^{-is\sigma^2} exp(-\frac{i}{4s}z_\mu C^{\mu\nu}z_\nu)\times \\
\nonumber
\biggl(\sigma+\frac{1}{2s}\gamma^\mu C_{\mu\nu}z^{\nu}-
\frac{e}{2}\gamma^\mu F_{\mu\nu}z^\nu\biggr)
\biggl[\tau \coth \tau-\frac{es}{2}\gamma^\mu\gamma^\nu F_{\mu\nu}\biggr].
\end{eqnarray}

Let us describe the  3D case to avoid some more complicated expressions.
The
EP is given by
\begin{equation} 
V_{eff}(\sigma)={\sigma^2\over 2\lambda}+ {1\over 4\pi^{3/2}}  
\int\limits_{1/\Lambda^2}^\infty {ds\over s^{5/2}}
e^{-s\sigma^2}(eBs) \coth(eBs)
\end{equation}

The most reliable method to keep all of the divergences is the cut-off
parameter introduction.
So we can make the following trick: write the integral in the EP like
\begin{equation}
\int\limits_{1/\Lambda^2}^\infty {ds\over s^{5/2}}
e^{-s\sigma^2}\left[ (eBs) \coth(eBs) - 1\right] + 
\int\limits_{1/\Lambda^2}^\infty {ds\over s^{5/2}}
e^{-s\sigma^2}
\end{equation}
and calculate the last one keeping
$\Lambda$ finite while the first integral is finite already and we can 
put $1/\Lambda^2=0$. Then it appears to be possible to calculate it as  a
 limit $\mu \to -1/2$ using the formula 
\begin{equation} 
\int\limits_0^\infty dx x^{\mu - 1}e^{-ax}\coth (cx)
= \Gamma (\mu) \left[ 2^{1 - \mu} 
(c)^{-\mu}\zeta(\mu , \frac{a}{2c})-a^{-\mu}\right].
\end{equation}

Finally the EP has the form
\begin{equation}
V_{eff}(\sigma)={\sigma^2\over 2\lambda}-
\left[ \frac{\Lambda \sigma^2}{2\pi^{3/2}}+
\frac{\sqrt{2}}{\pi} (eB)^{3/2} 
\zeta \left( -\frac{1}{2}, 1+ \frac{\sigma^2}{2eB} \right)+
\frac{1}{2\pi}eB \sigma \right]
\end{equation}

There are two ways of justifying the introduction of the $\Lambda$ parameter
in the formula for the EP.
The first one is the standard renormalization procedure, by means of
 the  UV cut-off
method. Then, in the limit $\Lambda\to\infty$, after  renormalization of
the coupling constant
\begin{equation}
\frac{1}{\lambda_R}=\frac{1}{\lambda}-\frac{\Lambda}{\pi^{3/2}},
\end{equation}
we have the   expression for the
 renormalized EP in  3D spacetime 
\begin{equation}
V_{eff}^{ren}(\sigma)={\sigma^2\over 2\lambda_R}-
-\frac{\sqrt{2}}{\pi} (eB)^{3/2} 
\zeta \left( -\frac{1}{2}, 1+ \frac{\sigma^2}{2eB} \right)
- \frac{1}{2\pi}eB \sigma
\end{equation}

For $B=0$ dynamical symmery breaking takes place when 
\begin{equation}
\lambda > \lambda_c=\frac{\pi^{3/2}}{\Lambda} 
\end{equation}  
if only we keep the finite cut-off $\Lambda$ 
meanwhile the renormalized NJL model does not admit 
this phenomenon in general.
However any finite value of the external magnetic field changes the situation
dramatically and dynamical symmetry breaking occurs for any
coupling constant. 
For $\sigma^2\ll eB$ nontrivial solution of the gap equation defining
a nontrivial minimum of the EP is given by 
\begin{equation}
\sigma=\frac{eB\lambda_R}{2\pi}
\end{equation}

The same calculations have been done for
a constant electric field. The nonzero imaginary 
part appears in this case caused by vacuum instability of the quantum field 
theory in 
electric field. But treating the real part of the EP we can 
find that 
electrical field restores the chiral symmetry, initially broken for the
finite cut-off parameter case.

Fig.1 illustrates the  universal character of magnetic catalysis.
It is a  plot of 3D $V_{eff}^{ren}(\sigma)$ with
$\mu=100;$
   $\lambda\mu=100$. Starting from above the curves correspond to  the 
following electromagnetic field 
configurations: $eE/\mu^2 = 0.0002, B=0$; $B=E=0$; $eB\mu^2=0.0002, E=0$.
 
After 
 renormalization the chiral
symmetry exists without external field but the magnetic field creates the 
non-zero minimum that indicates that DSB takes place. Meanwhile the external 
electric field works evidently against symmetry breaking.
In all figures,
an arbitrary dimensional parameter, $\mu$, defining a typical scale in
the model, is introduced in order to perform the plots in terms of
 dimensionless variables.

\begin{figure}
\psfig{figure=ESH_FIG_1,width=5.5in,height=3.5in}
\begin{center}
FIGURE 1
\end{center}
\end{figure}
\section*{General expression for effective potential in external 
electromagnetic and gravitational fields}
We have just the same expression for 
the EP in curved spacetime 
\begin{equation}
V_{eff}'(\sigma)={ \sigma \over \lambda }-i \Sp G(x,x,\sigma)
\end{equation}
To calculate the linear curvature corrections the local momentum   
expansion formalism is the most convinient one. Then in the
special Riemannian normal coordinate framework
\begin{equation}
g_{\mu\nu}(x)=\eta_{\mu\nu}-{1\over 3 } R_{\mu\rho\sigma\nu}y^\rho
y^\sigma
\end{equation}
with corresponding formulae for the others values and $y=x-x'$.

Then choosing the vector potential of the external electromagnetic field 
in the form 
\begin{equation}
A_\mu(x)=-{1 \over 2 } F_{\mu\nu} x^\nu ,
\end{equation}
where $F_{\mu\nu}$ is the constant matrix of electromagnetic field
strength tensor we find that:
\begin{equation}
G(x, x', \sigma)= 
\Phi (x, x')\left[ \tilde {G}_0 (x - x', \sigma)+
\tilde {G}_1 (x - x', \sigma) \dots \right],
\end{equation}
where $G_n \sim R^n $. 

Therefore  we obtain the iterative sequence
of equations for the  GF and the linear- curvature corrections are
given by
\begin{eqnarray}
\tilde{G}_1(x-x',\sigma)=\int\! dx''G_{00}(x-x'',\sigma)\times\\
\left[ -{i \over 6}
\gamma^a R^\mu{}_{\!\rho\sigma a}(x''-x')^\rho
(x''-x')^\sigma\partial_\mu^{x''}
\tilde{G}_0(x''-x',\sigma)-\right.\\
\nonumber
\left.{i \over 4}\gamma^a
\sigma^{bc}R_{bca\lambda}(x''-x')^\lambda\right]
\tilde{G}_0(x''-x',\sigma)
\nonumber
\end{eqnarray}
Here $G_{00}(x-x',\sigma)$ is a  free fermion GF. 

Substituting an exact flat spacetime GF of fermions in external
electromagnetic field  into this formula  
after some algebra we have evident expression for the
EP  with the linear-~curvature accuracy
in the constant curvature  spacetime.

\subsection*{External constant magnetic field case}

For 3D spacetime the EP  is given by
\begin{eqnarray}
V_{eff}(\sigma)={\sigma^2\over 2\lambda}+{1\over 4\pi^{3/2}}
\int_{1/\Lambda^2}^\infty {ds\over s^{5/2}}\exp(-s\sigma^2)\tau
\coth\tau-\\
{R\over 144\pi^{3/2}}\int_{1/\Lambda^2}^\infty
\int_{1/\Lambda^2}^\infty
\frac{ds dt}{(t+s)^{5/2}(1+\kappa
\coth\tau)^2}\exp[-(t+s)\sigma^2]\times\\
\nonumber
\biggl[2\kappa(\kappa+\tau)+
(9\tau+5\kappa)\coth\tau+\kappa(\tau-3\kappa)\coth^2\tau\biggr]
\end{eqnarray}
where $\tau=eBs, \kappa= eBt$.

We can perform the same renormalization procedure as in flat spacetime 
because no 
new divergences appear in the linear-curvature corrections. But we keep the 
cut of scheme here to study the most general situation.

The results are presented on  Fig.2.
It shows a plot of 3D $V_{eff}^{ren}(\sigma)$ with
$\mu=100;
    \lambda\mu=100$ and fixed $eB\mu^2= 0.0002$.
Starting from above the
curves correspond to  the different values of spacetime curvature
$R\mu^2=  0.0025,  0.002, 0.001, 0$.
 Second-order phase transition 
ruled by the spacetime curvature takes place. 

\begin{figure}
\psfig{figure=ESH_FIG_2,width=5.5in,height=3.5in}
\begin{center}
FIGURE 2
\end{center}
\end{figure}

\subsection*{External constant electrical field case}

We have for renormalized EP the following expression
\begin{eqnarray}
V_{eff}^{ren}(\sigma)={\sigma^2\over 2\lambda_R}-
{(2ieE)^{3/2} \over 4\pi}
\biggl[ 2 \zeta (-{1\over2},{\sigma^2 \over 2ieE})-
\biggl({\sigma^2\over 2ieE}\biggr)^{1/2} \biggr]+\\
{R\sigma \over 24\pi}+{iR(eE)^{1/6} \over 2\pi^2 3^{7/3}}
\exp(-\pi{\sigma^2 \over eE})\Gamma(\frac{2}{3})\sigma^{2/3}.
\nonumber
\end{eqnarray}

Here we have performed a small electric field expansion in the 
 R-dependent term.
A numerical analysis 
of $\Re V_{eff}(\sigma)$
for negative coupling constant gives the typical behaviour of a
first--order phase transition, as shown
in Fig. 3. The critical values are defined as usual:
$R_{c1}$ corresponds to the spacetime curvature  for which
a local nonzero minimum appears,  $R_{c}$, when the real part of 
EP is equal at zero and at the local minimum, and  $R_{c2}$, when the zero
extremum becomes a maximum. 
There is a plot   of 3D $\Re V_{eff}^{ren}/\mu^3$
as a function
of $\sigma/\mu$  for
fixed $eE/\mu^2= 0.00005$ and  $ \lambda\mu= -100$.
From above to below,  the curves in the  plot correspond to
the following values of
$R/\mu^2=0.006; 0.005; 0.004; 0.0032; 0$, respectively.
The critical values, defined as usual, are given by:
$R_{c1}/\mu^2=0.005$; $R_{c}/\mu^2=0.0032$;
 $R_{c2}/\mu^2=0$.
$\Lambda$ obviously does not appear anywhere because after renormalization
it must be sent to infinity, $\Lambda\to\infty$.
\begin{figure}
\psfig{figure=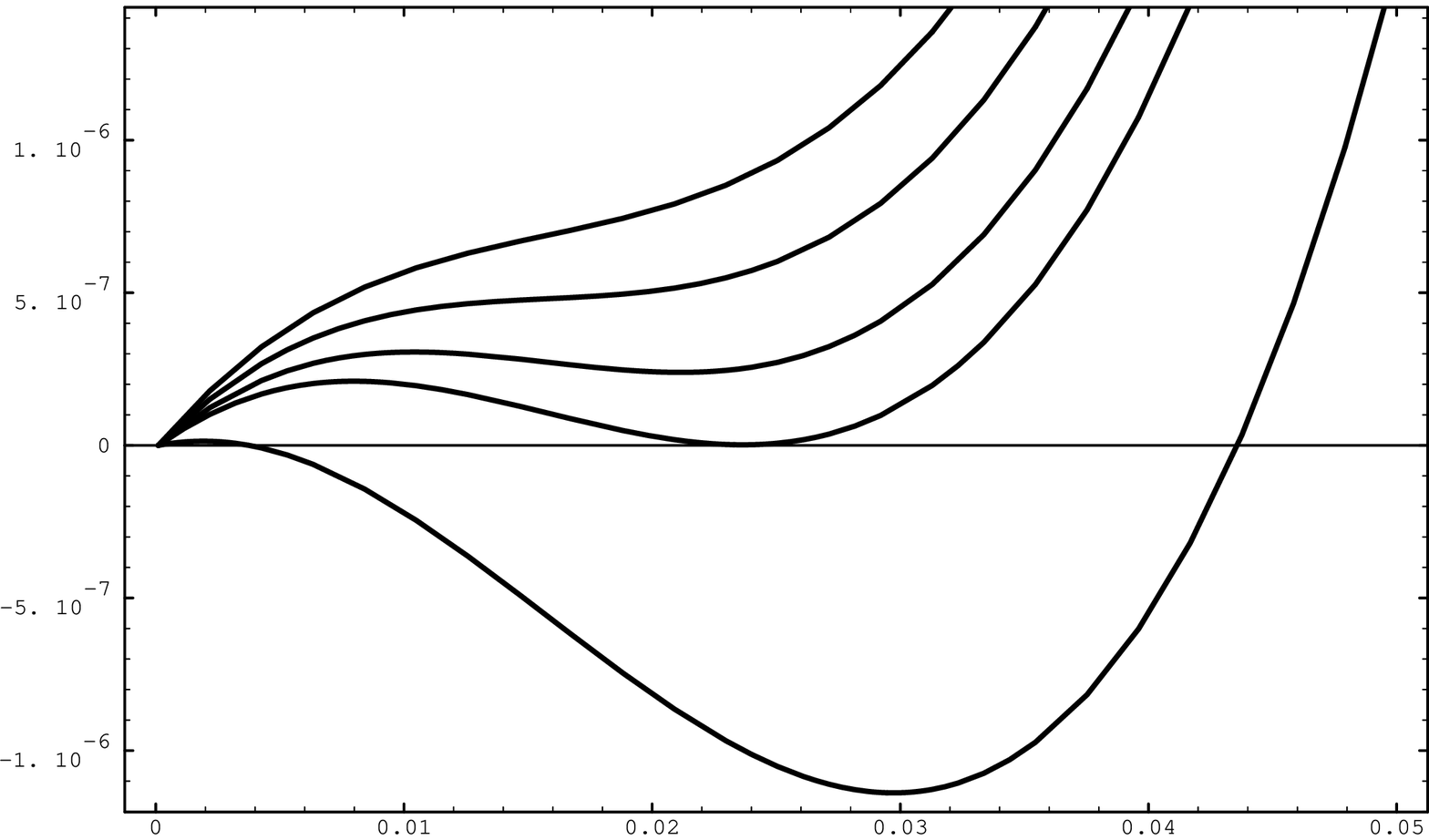,width=5.5in,height=3.5in}
\begin{center}
FIGURE 3
\end{center}
\end{figure}

\section*{Conclusions} 

We clearly observe that a positive spacetime curvature tries
to restore chiral symmetry even in the presence of external magnetic field.
Therefore the universal character of magnetic catalysis doesn't
survive in curved spacetime. From the other hand electric field 
increases the critical value of coupling constant as it does in flat spacetime.

It should be noted that for $D<4$ is renormalizable and these conclusions don't 
depend already on the cut-off scale $\Lambda$. 

This work has been partly financed by DGICYT (Spain), project PB96-0095, and
by  CIRIT (Generalitat de Catalunya),  grant 1995SGR-00602.
The work of Yu.I.Sh.  was supported in part by Ministerio de
Educaci\'on y Cultura (Spain), grant SB96-AN4620572.

\end{document}